\documentclass[twocolumn,showpacs,preprintnumbers,amsmath,amssymb]{revtex4}
\usepackage{graphicx}
\usepackage{bm}

\begin{document}

\title{Capillary flow in an evaporating sessile drop}
\author{Yuri Yu. Tarasevich}
\affiliation{Astrakhan State University, Tatishcheva 20à,
Astrakhan, 414056, Russia}
 \email{tarasevich@astranet.ru}


\begin{abstract}
An analytical expression of velocity potential inside an
evaporating sessile drop with pinned contact line is found.
\end{abstract}

\maketitle

\section{\label{sec:int}Introduction}
The desiccated sessile drops attract the attention because of
different reasons and possible applications. If a drop of pure
liquid dries on a smooth substrate, then the base of the drop
shrinks but the contact angle remains fixed. Under evaporation, a
drop of colloidal suspension or solution having a strongly
anchored three phase line keeps a spherical-cap shape with a
constant base. The contact angle decreases with time and the
height in each point of the profile decreases. To satisfy the
anchoring condition, a flow of liquid has to occur inside the
drop.  An outward flow in a drying drop is produced when the
contact line is pinned so that liquid that is removed by
evaporation from the edge of the drop must be replenished by a
flow of liquid from the interior
\cite{Parisse96,Deegan1997,Deegan2000a}. This flow is capable of
transferring 100\% of the solute to the contact line and thus
accounts for the strong perimeter concentration of many stains.
Ring formation in an evaporating sessile drop is a hydrodynamic
process in which solids dispersed in the drop are advected to the
contact line. After all the liquid evaporates, a ring-shaped
deposit is left on the substrate that contains almost all the
solute. Perhaps, everybody knows a dense, ring-like deposit along
the perimeter of a dried drop of coffee or tee, milk or juice on a
table.

Exploratory experiments \cite{Deegan2000a} using a variety of
carrier fluids, solutes, and substrates indicated that
preferential deposition at the contact line is insensitive to a
wide range of experimental conditions. Ringlike deposits were
observed whenever the surface was partially wet by the fluid
irrespective of the chemical composition of the substrate.
Different substrates were investigated (metal, polyethylene,
roughened Teflon, freshly cleaved mica, ceramic, and silicon).
Rings were found in big drops (15 cm) and in small drops (1 mm).
They were found with aqueous and non-aqueous (acetone, methanol,
toluene, and ethanol) solvents. They were found with solutes
ranging in size from the molecular (sugar and dye molecules) to
the colloidal (10 $\mu$m polystyrene microspheres) and with solute
volume fractions ranging from 10$^{-6}$ to 10$^{-1}$. Likewise,
environmental conditions such as temperature, humidity, and
pressure could be extensively varied without affecting the ring.
Effects due to solute diffusion, gravity, electrostatic fields,
and surface tension forces are negligible in ring formation.


The vertically averaged radial flow of the fluid in a desiccated
sessile drop was calculated in the work \cite{Parisse96}. The
conservation of fluid was utilized. A constant evaporation rate
all over drop free surface was assumed.

For the case where the limiting rate is the diffusion of the
liquid vapor, the evaporation of the drop rapidly attains a steady
state so that the diffusion equation reduces to Laplace's equation
\cite{Deegan2000a}. In this case the evaporation rate is larger
near the contact line and the resulting ringlike deposit is more
concentrated at the edge.

The models \cite{Parisse96} and \cite{Deegan2000a} deal with the
vertically averaged radial flow of the fluid. The next Section
describes how to obtain the space distribution of the fluid flow.

\section{\label{sec:model}Model and results}
Likewise the works \cite{Parisse96,Deegan1997,Deegan2000a}, let us
suppose, that the shape of the drop is a spherical cap. It means
that the drop is small and a spherical cap shape induced by
surface tension (Bond number smaller than one: $\mathrm{Bo} =
g(\rho - \rho_f)d^2 \sigma^{-1} \ll 1$, where $g$ is gravitational
acceleration, $\rho$ is drop density, $\rho_f$ is surrounding
medium density , $\sigma$ is surface tension, $d$ is drop
diameter). The problem of a spherical cap on an impermeable
substrate can be replaced by the problem of a lens. Let us
consider an extremely simple situation when the cap is a
semi-sphere, i.e. the contact angle is 90$^\circ$.

Under evaporation, the height in each point of the profile
decreases. Nevertheless, under room condition the evaporation is a
slow process. So, the desiccation time of a 15~mg sessile water
drop is larger then 3500~s \cite{Annarelli2001}. This fact gives a
possibility to consider a quasi-static process.  We will consider
a drop with a fixed free surface and the specific boundary
conditions. Quasi-static approximation was utilized in
\cite{Deegan2000a} to obtain a vapor rate near the free surface,
too.

Let us suppose that the evaporation rate is uniform all over drop
free surface, i.e. we will consider only begin of the evaporation.
The conservation of fluid determines the relationship between the
velocity of the free surface, $u_n$, the normal to the free
surface flow of the fluid, $v_n$, and the rate of mass loss per
unit surface area per unit time from the drop by evaporation, $J$:
$$\rho v_n + J = \rho u_n.$$

Let us consider the potential flow, only. The potential of the
velocity field, $\varphi$, is a solution of the Laplace's equation

\begin{equation}\label{eq:Laplace}
  \Delta \varphi = 0.
\end{equation}
Also, we will solve the Laplace's equation inside a spherical area
using spherical coordinates .

Measurements of the evolution of the drying drop height $h$ with
time $t$ can be performed. Let us suppose, that the velocity of
the drop height, $u_0$, is known. Thus $u_n = u_0 \cos \theta,$
where $\theta$ is azimuthal angle.  The volume of the drop is
decreasing due to evaporation: $$\int (J - \rho u_0 \cos \theta)\,
dS = 0,$$ where the integration goes all over the free surface.
Thus $$J = \frac {u_0 \rho}{2}.$$ The equation of solute
conservation yields,
\begin{equation}\label{eq:vr}
 v_r(r = R) = \left.\frac{\partial \varphi}{\partial r}\right|_{r
 = R} = u_0\left(\frac{1}{2} - |\cos \vartheta| \right),
\end{equation}
where $R$ is the radius of the contact base.

We looking for the boundary value problem \eqref{eq:Laplace},
\eqref{eq:vr} inside a spherical area $(r \leqslant R)$ as
\begin{equation}\label{eq:sol}
\varphi(r,\vartheta) = \sum_{k = 0}^\infty A_k\left(
\frac{r}R{}\right)^k P_k(\cos \vartheta),
\end{equation}
where $P_k(\cos \vartheta)$ are  Legendre polynomials.

Thus,
\begin{equation}\label{eq:dphidr}
  \frac{\partial \varphi}{\partial r} = \sum_{k = 0}^\infty k
  \frac{A_k}{R^k}r^{k - 1}P_k(\cos \vartheta).
\end{equation}

Let us write the right hand side of the equation \eqref{eq:vr}
using Fourier--Legendre series expansion also known as a
Generalized Fourier Series expansion:
\begin{equation}\label{eq:vrser}
  u_0 \left( \frac{1}{2} - |\cos \vartheta| \right) = \sum_{k =
  0}^\infty b_k P_k (\cos \vartheta).
\end{equation}

Where, $$
  b_k = \frac{2k + 1}{2} u_0 \int\limits_{0}^{\pi} \left( \frac{1}{2} -
  |\cos \vartheta| \right)P_k (\cos \vartheta)\sin \vartheta
  \, d\vartheta.
$$

Thus, $b_0 = 0$, $b_{2k + 1} = 0$, $$b_{2k} = u_0 (4k + 1)
\frac{(-1)^k (2k - 2)!}{2^{2k}(k - 1)!(k + 1)!}, \qquad
k=1,2,\dots . $$

Taking into account \eqref{eq:vr},  \eqref{eq:dphidr} equals
\eqref{eq:vrser}, if $r = R$:
\begin{equation}\label{eq:bk}
 \sum_{k = 0}^\infty k
  \frac{A_k}{R^k}R^{k - 1}P_k(\cos \vartheta) = \sum_{k =
  0}^\infty b_k P_k (\cos \vartheta).
\end{equation}

Eq. \ref{eq:bk} is always valid, if $A_k = b_k R/k$. Hence, the
coefficients of the series \eqref{eq:sol} are
\begin{equation*}\label{eq:ak}
  A_{2k} = u_0 R (4k + 1)
\frac{(-1)^k (2k - 2)!}{2^{2k}k!(k + 1)!}, \qquad k=1,2,\dots .
\end{equation*}

\begin{figure}[!htbp]
  \centering
  \includegraphics*[width=0.9\linewidth]{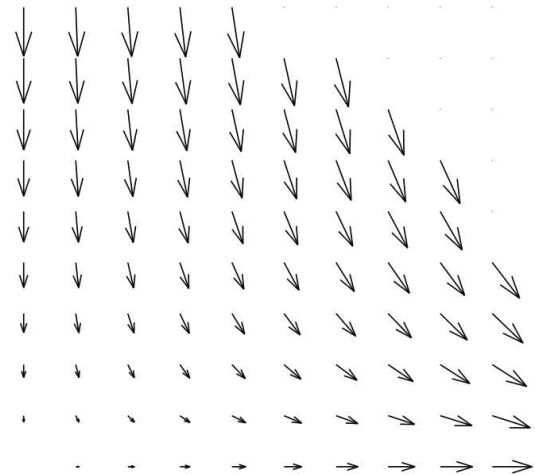}
  \caption{Capillary flow in a desiccated sessile drop
  }\label{fig:flow}
\end{figure}

The capillary flow carries a fluid from the apex of the drop to
the contact line (Fig.~\ref{fig:flow}).

\section{\label{sec:concl}Discussion}
It is clear, more realistic assumption, that a drop has a
spherical cap shape with an arbitrary contact angle, can not
change the qualitative picture of the flow.  In particularly, the
calculations of the vertically averaged velocity for the drops
with the contact angles up to 90$^\circ$ \cite{Parisse96} did not
show the qualitative changes of the flow. If a drop has a
spherical cap shape with an arbitrary contact angle the Laplace's
equation can be solved using toroidal coordinates. Although, the
analytical results are too complicate in this case, and numerical
calculations should be performed. Nonuniform evaporation rate can
not change the results qualitative, but the line, where
$v_r(R)=0$, has to move to the base.

Recently, the buckling instability was investigated during the
drying of sessile drops of colloidal suspensions or of polymer
solutions \cite{Pauchard2003,Pauchard2003a,Pauchard2003b}. Under
solvent evaporation, disperse particles or polymers accumulate
near the vapor--drop interface. The outer layer of the drop is
more concentrated in the polymer and may display a gel or glassy
transition and, hence, may form a permeable rigid gelled or glassy
skin. This skin behaves like an elastic shell although it does not
block the evaporation. This gelled or glassy skin will thus bend
as the volume it encloses decreases, leading to large surface
distortions.

The processes of skin formation are not the scope of this brief
communication. Nevertheless, knowledge of the velocity field
inside an evaporating sessile drop is a necessary background to
describe the colloidal particles transfer or solute diffusion, and
skin formation.


\begin{thebibliography}{9}

\bibitem{Parisse96}
F. Parise and C. Allain,  J. Phys. II France  \textbf{6}, 1111
(1996).

\bibitem{Deegan1997}
R. D. Deegan, O. Bakajin, T. F. Dupont, G. Huber, S. R. Nagel, and
T. A. Witten, Nature (London) 389, 827 (1997).

\bibitem{Deegan2000a}
R. D. Deegan, O. Bakajin, T. F. Dupont, G. Huber, S. R. Nagel, and
T. A. Witten, Phys. Rev. E.  \textbf{62}, 756 (2000).

\bibitem{Annarelli2001}
C. C. Annarelli, J. Fornazero, J. Bert, and J. Colombania, Eur.
Phys. J. E  \textbf{5}, 599 (2001).

\bibitem{Pauchard2003}
L. Pauchard and C. Allain,  Europhys. Lett. {\bf 62}(6), 897 (
2003).

\bibitem{Pauchard2003a}
L. Pauchard and C. Allain,  C. R. Physique  {\bf 4}, 231 (2003).

\bibitem{Pauchard2003b}
L. Pauchard and C. Allain,  Phys. Rev. E  {\bf 68}, 052801 (2003).


\end{thebibliography}
\end{document}